\documentclass[preprint,showpacs,preprintnumbers,superscriptaddress]{revtex4-1}
\usepackage{newunicodechar}
\newunicodechar{Λ}{\Lambda}
\usepackage{multirow}
\usepackage{diagbox}
\usepackage[colorlinks=true,          
            linkcolor=red,          
            citecolor=blue,         
            urlcolor=blue]{hyperref} 
\usepackage{booktabs}
\usepackage{array}
\usepackage{natbib}
\usepackage{amsmath,amssymb,graphicx,bm,subfigure,float,siunitx,color}
\usepackage{bm,subfigure,float,siunitx,graphicx}
\usepackage[figuresright]{rotating}
\usepackage{color}
\usepackage{epstopdf}
\usepackage{mathrsfs}

\begin{document}
\title{A parameterized equation of state for dark energy and Hubble Tension}

\author{Jing-Ya Zhao}
\affiliation{College of Physics Science and Technology, Hebei University, Baoding 071002, China}
\author{Tong-Yu He}
\affiliation{International Center of Supernovae (ICESUN),Yunnan Key Laboratory of Supernova Research,Yunnan Observatories, Chinese Academy of Sciences, Kunming 650216, China}
\affiliation{Key Laboratory for the Structure and Evolution of Celestial Objects, CAS, Kunming 650216, P.R. China}
\author{Jia-Jun Yin}
\affiliation{College of Physics Science and Technology, Hebei University, Baoding 071002, China}
\author{Zhan-Wen Han}
\affiliation{International Center of Supernovae (ICESUN),Yunnan Key Laboratory of Supernova Research,Yunnan Observatories, Chinese Academy of Sciences, Kunming 650216, China}
\affiliation{Key Laboratory for the Structure and Evolution of Celestial Objects, CAS, Kunming 650216, P.R. China}
\affiliation{University of Chinese Academy of Sciences, Beijing 100049, China}
\author{Rong-Jia Yang \footnote{Corresponding author}}
\email{yangrongjia@tsinghua.org.cn}
\affiliation{College of Physics Science and Technology, Hebei University, Baoding 071002, China}
\affiliation{Hebei Key Lab of Optic-Electronic Information and Materials, Hebei University, Baoding 071002, China}
\affiliation{National-Local Joint Engineering Laboratory of New Energy Photoelectric Devices, Hebei University, Baoding 071002, China}
\affiliation{Key Laboratory of High-pricision Computation and Application of Quantum Field Theory of Hebei Province, Hebei University, Baoding 071002, China}

\begin{abstract}

We propose a parameterized equation of state for dark energy and perform observational tests with the Hubble parameter measurements, the Pantheon supernova sample, and DESI DR2 data. We obtain the best-fit values for the parameters as: $H_0 = 73.28\pm 0.15~\mathrm{km\,s^{-1}\, Mpc^{-1}}$, $\Omega_{\rm m} = 0.316\pm 0.011$, and $\alpha = 0.00032\pm 0.00046$, demonstrating that the model exhibits a high degree of consistency with astronomical observations and provides a promising parameterized method for addressing the Hubble tension.


\end{abstract}

\maketitle

\section{Introduction}

One of the most important discoveries in modern cosmology is that the Universe is currently undergoing an accelerated expansion, which poses a challenge to general relativity as a complete theory of gravity. Under the assumption that general relativity remains valid on cosmological scales, this phenomenon requires the introduction of an unknown energy component, commonly referred to as dark energy. The $\Lambda$CDM framework continues to be the leading and most widely supported description of dark energy. However, it suffers from several long-standing theoretical difficulties, including the cosmological constant problem \cite{Carroll:2000fy}, characterized by the enormous discrepancy between theoretical expectations and observational measurements; the coincidence problem \cite{Zlatev:1998tr}, which asks why the energy densities of dark energy and matter are of the same order of magnitude at the present epoch; and the so-called age problem discussed in certain observational contexts \cite{Yang:2009ae}. Moreover, a pronounced discrepancy persists between determinations of the Hubble constant derived from early- and late-universe observations.

In the contemporary cosmological framework, the Hubble constant $H_{0}$, which characterizes the present expansion rate of the universe, plays a central role in probing the fundamental properties of both dark energy and dark matter. Current determinations of $H_{0}$ rely mainly on two independent approaches: indirect constraints from early universe probes \cite{Durrer:2013pga} and direct measurements from late universe observations \cite{daCosta:2023mow}. Early Universe measurements infer the value of the Hubble constant through precision observations of cosmic microwave background (CMB) anisotropies within the $\Lambda$CDM framework. Based on the final Planck data release, the most precise early Universe determination yields $H_{0} = 67.4 \pm 0.5~\mathrm{km\,s^{-1}\,Mpc^{-1}}$ \cite{Planck:2018vyg}. In contrast, late time measurements based on the combined Cepheid and supernova samples give a significantly higher value, $H_{0} = 73.04 \pm 1.04~\mathrm{km\,s^{-1}\,Mpc^{-1}}$ \cite{Riess:2021jrx}. The discrepancy between these two determinations reaches the level of approximately $5\sigma$, well beyond the range expected from statistical uncertainties. This tension is now widely known as the Hubble tension \cite{DiValentino:2020zio, Wang:2023bxf, Verde:2019ivm, Abdalla:2022yfr, CosmoVerseNetwork:2025alb}. Related aspects of this problem have been investigated in Refs.~\cite{Dainotti:2025qxz, He:2024jku, Paul:2025wix, Milne:2014rfa}. In addition to the discrepancy between early and late Universe measurements, indications of tension have also been reported in certain low-redshift observational data. Despite extensive efforts, including modifications of early Universe physics \cite{Niedermann:2020dwg}, the introduction of new physical components \cite{Gogoi:2020qif}, and detailed reassessments of potential systematic effects, the Hubble tension persists as one of the most striking challenges in modern cosmology. 

Since the Hubble tension is primarily established within the $\Lambda$CDM framework, it may signal the need for dynamical dark energy beyond a simple cosmological constant. Analysis of observational Hubble data or Dark Energy Spectroscopic Instrument (DESI) DR2 data with a model-independent method implies a dynamical dark energy \cite{Yang:2023qsz,Yang:2025cgx,Yang:2025kjq}. Recent observations from the DESI DR2 data, combined with independent measurements of the Hubble parameter, also indicate a statistical preference for dynamical dark energy models over the standard $\Lambda$CDM scenario \cite{DESI:2024mwx,  Gadbail:2024lek, Samaddar:2025xak, 2025NatAs.tmp..204G}.
These developments provide strong motivation to investigate concrete realizations of dynamical dark energy. Here we propose a parameterized equation of state (EoS) for dark energy and constrain the parameters with observational data from Hubble parameter measurements, the Pantheon supernova sample, and DESI DR2 data. The value of $H_0$ we obtain is consistent with that obtained from the combined Cepheid and supernova samples \cite{Riess:2021jrx}, which can alleviate the Hubble tension.

In  \cite{Yang:2009zzl}, an EoS for dark energy was obtained from the k-essence model. Here, we generalize this EoS so that it can evolve like quintessence, cosmological constant, or phantom dark energy. We aim to address two specific questions: (i) Can this parameterization provide a good fit to current cosmological observations? (ii) Can this model alleviate the approximately $5\sigma$ Hubble tension between early-universe (Planck) and late-universe (SH0ES) measurements? 

The structure of this paper is as follows. Sec. II provides a concise overview of the k-essence cosmology considered in this work. In Sec. III, we employ Markov Chain Monte Carlo (MCMC) techniques to constrain the model parameters using current observational datasets. Sec. IV presents a discussion of the resulting constraints and Sec. V summarizes our main conclusions.

\section{The parameterized equation of state}

There are generally two parameterization methods for dark energy cosmology: one is to parameterize the Hubble function, and the other is to parameterize the EoS of dark energy. In \cite{Yang:2009zzl}, an EoS for dark energy was derived from a class of k-essence model. Here we generalize this EoS for dark energy so that it can behave like quintessence, cosmological constant, or phantom energy 
\begin{eqnarray} 
\label{14}
w_{\rm x}=-\frac{1}{1+\alpha(1+z)^6},
\end{eqnarray}
where $\alpha$ is a positive constant in \cite{Yang:2009zzl}. Here we relax this restriction and treat $\alpha$ as a general constant: when $\alpha=0$, the model reduces to $\Lambda$CDM; if $\alpha<0$, the model behaves like phantom dark energy; if $\alpha>0$, the model evolves like quintessence. Compared to $w_0w_a$CDM, Eq. \ref{14} has only one parameter, which helps to reduce the degeneracies between parameters.

With Eq. \ref{14}, the Friedmann equation in a flat Friedmann-Robertson-Walker-Lema\^{i}tre takes the form of 
\begin{eqnarray}
\label{15}
H^{2}(z)=H^2_0\left[\Omega_{\rm m}(1+z)^3+(1-\Omega_{\rm m})f(z)\right],
\end{eqnarray}
where $\Omega_{\rm m}$ is the present value of dimensionless energy density for dark and baryonic matter, and
\begin{eqnarray}
\label{16}
f(z)=\sqrt{\frac{1+\alpha(1+z)^6}{1+\alpha}}.
\end{eqnarray}
If $\alpha<0$, the EoS in Eq. \ref{14} will be divergent at a certain redshift, but at the same time we have $f(z)=0$ in Eq. \ref{16} and $H(z)$ is still regular, meaning that during periods of higher redshift, the influence of dark energy disappears. In the model we consider, $H_0$, $\Omega_{\rm m}$, and $\alpha$ are the parameters needed to be constrained with observational data, which is exactly the main content of the next section.

\section{Methodology and data}

Building on the parameterized EoS for dark energy developed earlier, we next evaluate whether the approximate parameter values are capable of reproducing the observational data relevant to today's Universe. For this purpose, we primarily utilize three distinct datasets: the Hubble parameter dataset, the Pantheon sample of supernovae, and the DESI DR2 data. Before proceeding with the joint constraint of the model using the datasets, it is necessary to introduce the MCMC \cite{Rodriguez:2013oaa}: which operates on the core principle of constructing a specialized Markov chain based on the model. When the number of iterations becomes large enough, the resulting sample set converges to the posterior distribution of the model parameters. A notable feature of this method is that the convergence outcome is independent of the initial values.

The specific computational procedure is described as follows: First, we explicitly specify the parameters to be constrained ($H_{0}$, $\Omega_{\rm m}$, and $\alpha$) based on the cosmological model proposed in the preceding section, assigning physically motivated priors to each parameter and constructing the corresponding likelihood function. During the sampling process, we combine the three distinct data sets employing a joint likelihood function, which appropriately weights and combines the error distributions across the different observational datasets.
In this process, we utilize the \text{emcee} package \cite{Zuntz:2014csq} 
and present the results in the form of two-dimensional contour plots with confidence intervals of $68.3\%$ ($1\sigma$) and $95.5\%$ ($2\sigma$).

\subsection{Observational Hubble data}

One of the most critical data sets for constraining cosmological models is the observational Hubble data (OHD), which is obtained through direct measurements of the expansion rate\cite{Paxton:2019lxx,Mandal:2022ocg,Larison:2023kas}. In this study, we employ the differential age method to determine the Hubble parameter as a function of redshift \cite{Moresco:2018xdr,LozanoTorres:2024tnt},
\begin{equation}
H(z) = -\frac{1}{1+z}\frac{dz}{dt}
\label{eq:hubble}
\end{equation}
The redshift dependence of the Hubble parameter is determined using the differential age method. 
Also known as the cosmic chronometer approach, this technique computes the expansion rate at a given cosmic epoch by measuring the age difference ($dt$) between passively evolving galaxies formed at different redshifts, in conjunction with the corresponding redshift interval ($dz$). As shown in Eq. \ref{eq:hubble}, this enables a direct determination of the Hubble parameter without relying on assumptions from any specific cosmological model.

The cosmic chronometer approach provides a portion of $H(z)$ measurements derived from the cosmic expansion rate \cite{Zhang:2012mp,Stern:2009ep,Moresco:2012jh,Moresco:2015cya}, while another portion of $H(z)$ measurements is obtained from the analysis of BAO peaks in the galaxy power spectrum \cite{Gaztanaga:2008xz,Chuang:2012qt,Blake:2012pj,BOSS:2012gof,Oka:2013cba,Borghi:2021rft,eBOSS:2018yfg,BOSS:2016wmc,BOSS:2016zkm}, as collected in \cite{Yang:2023qsz}.
The cosmic chronometer method offers the distinctive advantage of being model-independent, whereas the BAO technique extends observational coverage over a broader redshift range. Owing to its model-independent nature, the cosmic chronometer method serves as a key diagnostic for probing dark energy models and addressing outstanding challenges such as the Hubble tension, thereby holding particular value in contemporary cosmological research.

To avoid overlapping data sets in BAO measurements, we only adopt the data derived from the cosmic chronometer method \cite{Myrzakulov:2023rxp}. The dataset spans redshifts from $z=0.07$ to $z=1.965$. For consistency, we adopt a representative value of $H_{0} = 70~\mathrm{km\,s^{-1}\,Mpc^{-1}}$ throughout the analysis~\cite{Perivolaropoulos:2021jda}. Subsequently, we performed a maximum likelihood analysis to determine the mean values of the model parameters ($H_{0}$, $\Omega_{\rm m}$ and $\alpha$).
The $\chi^2$ function\cite{Koussour:2026uyi} used for OHD data is given by
\begin{equation}
\label{9}
\chi^{2}_{\rm H(z)}=\Delta H^{T}C^{-1}_{\rm H(z)}\Delta H,
\end{equation}
where $\Delta H$ is the vector of residuals, with components
$\Delta H_i = H_{\mathrm{obs}}(z_i) - H_{\mathrm{th}}(z_i)$, and $C_{\rm H(z)}$ is the covariance matrix of the observational Hubble data. When the measurements are independent, $C_{\rm H(z)}$ is diagonal, with elements $(C_{\rm H(z)})_{ii} = \sigma^{2}_{H(z_i)}$ \cite{Moresco:2020fbm,Jia:2022ycc}.

\begin{figure}[htbp]
  \centering
  \includegraphics[width=0.6\textwidth]{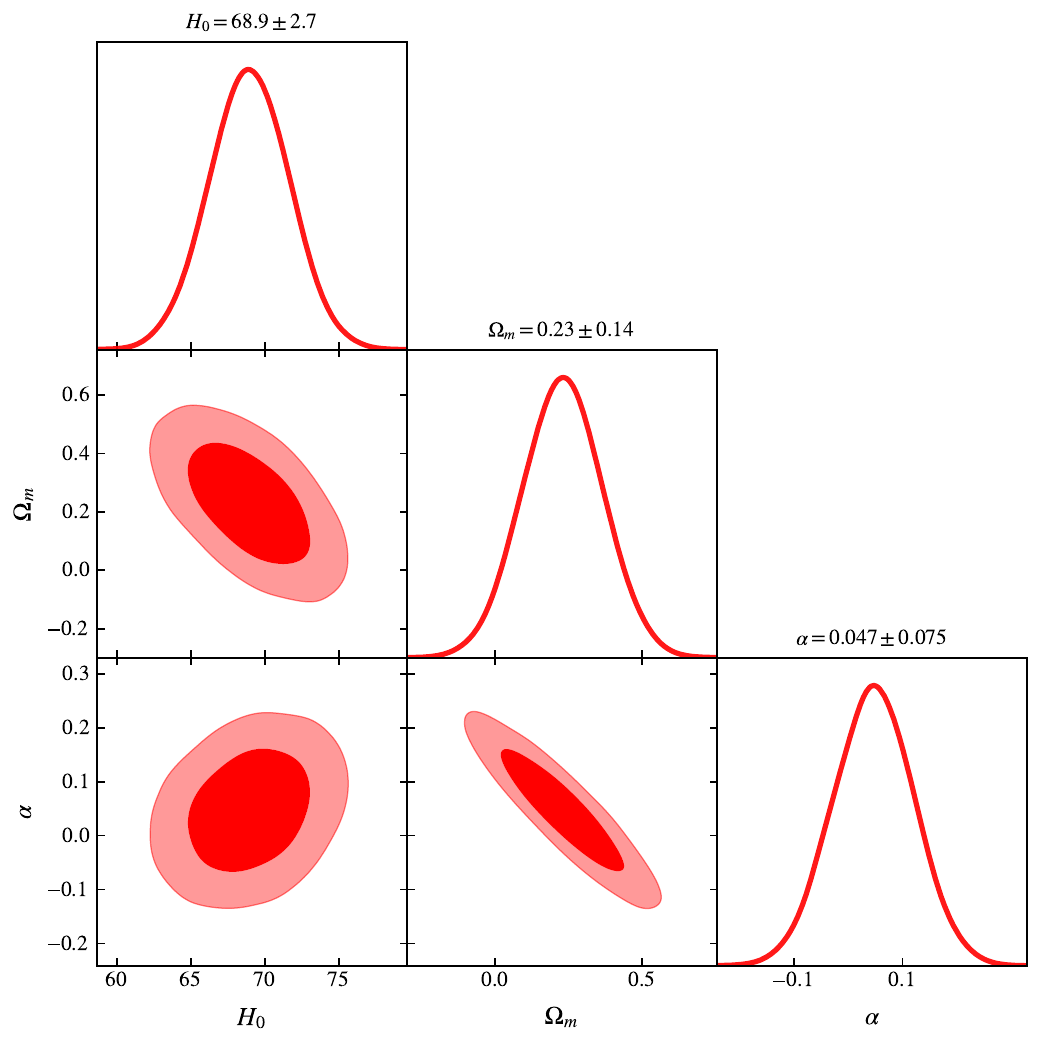}
  \caption{The inferred best-fit model parameters along with their \mbox{$1\sigma$} and 
\mbox{$2\sigma$} confidence regions based on the Hubble datasets.
  }
  \label{fig1}
\end{figure}

\begin{figure}[htbp]
  \centering
  \includegraphics[width=0.6\textwidth]{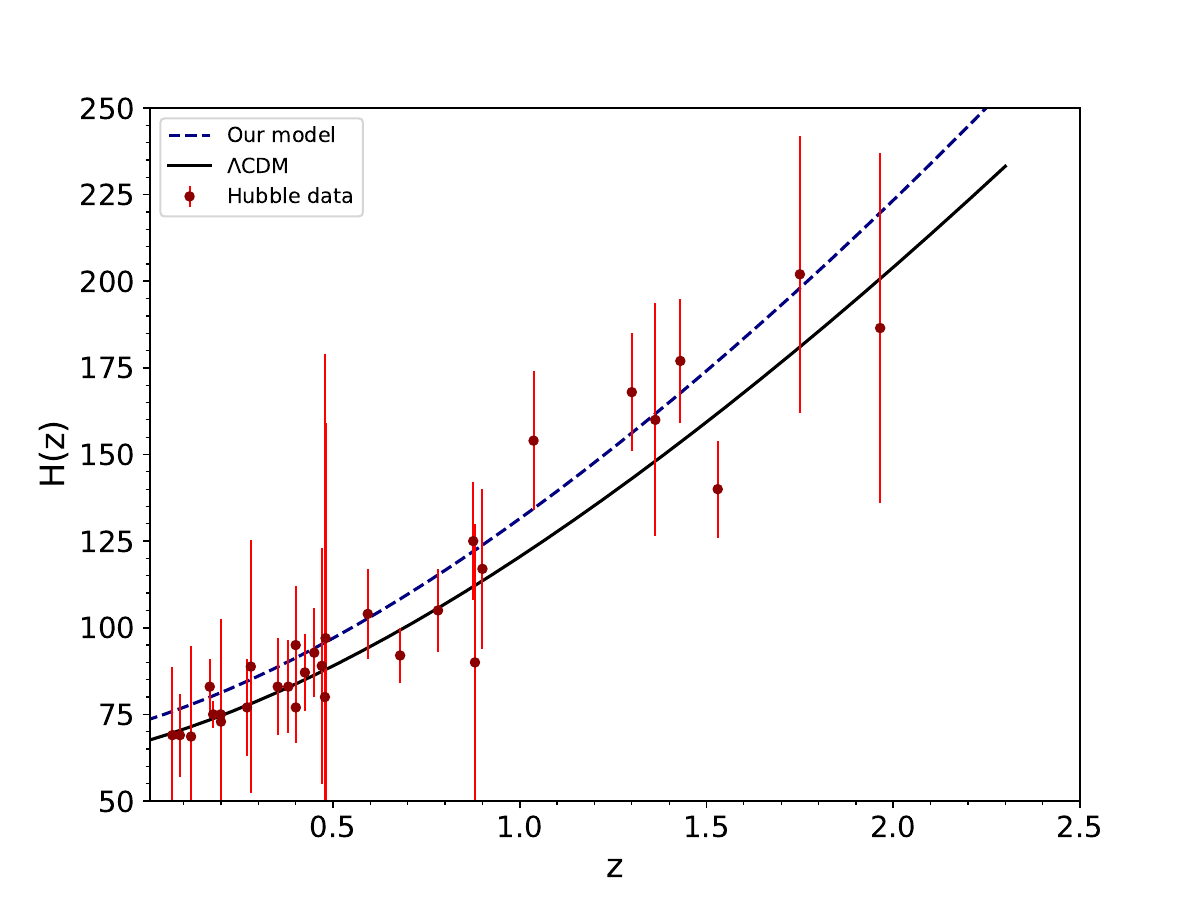}
  \caption{The reconstructed Hubble function $H(z)$ as a function of redshift inferred from observational Hubble data. The dotted lines show the prediction of our model, while the solid black curve denotes the $\Lambda$CDM scenario.
  }
  \label{fig2}
\end{figure}

\begin{figure}[htbp]
  \centering
  \includegraphics[width=0.6\textwidth]{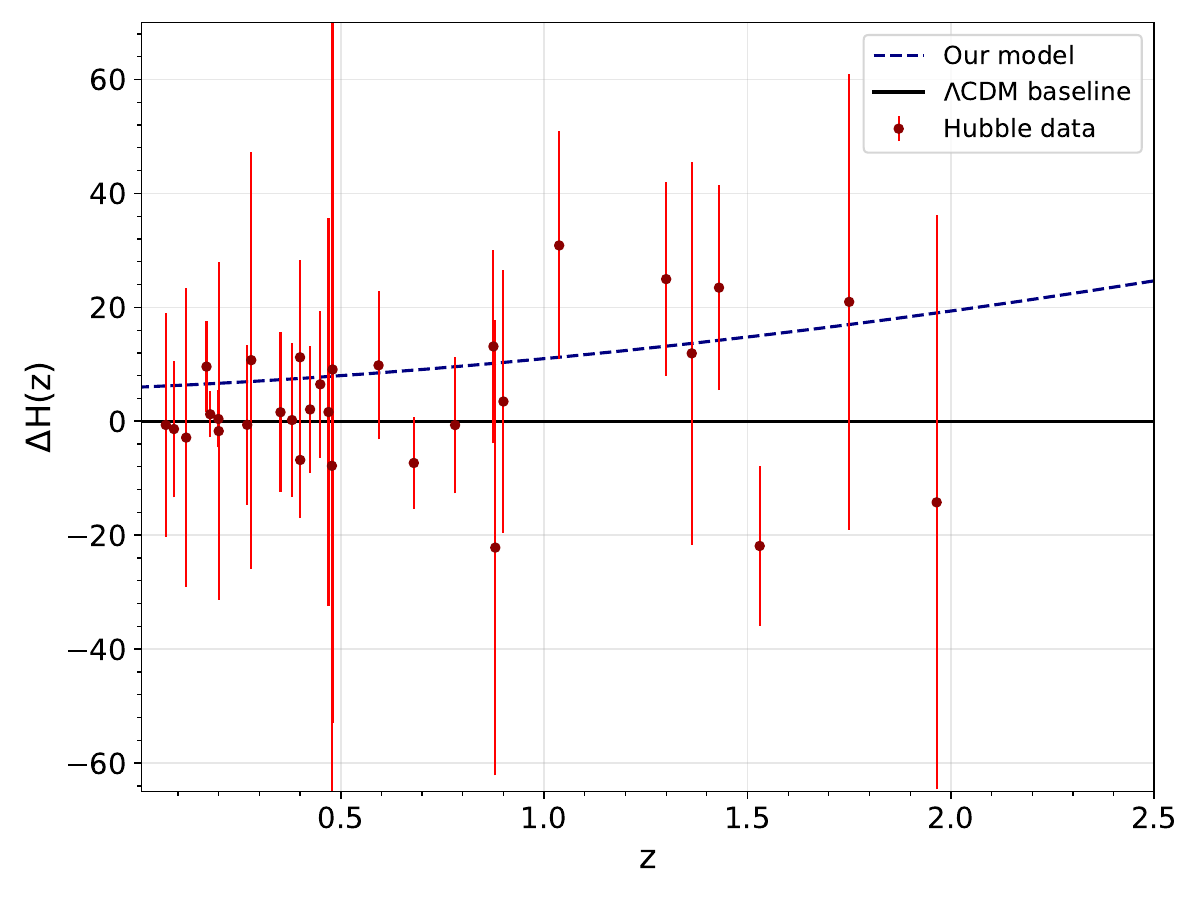}
  \caption{The residual plot comparing the Hubble function $H(z)$ reconstructed from observational Hubble data with the $\Lambda$CDM model. The dotted line shows the prediction of our model, while the solid black curve denotes the $\Lambda$CDM scenario.
  }
  \label{fig3}
\end{figure}

As shown in Figure~\ref{fig1}, we derived the best fit values of the model parameters ($H_0$, $\alpha$ and $\Omega_{\rm m}$) following the contours of the confidence levels of $68\%$ ($1\sigma$) and $95\%$ ($2\sigma$).
At the $1\sigma$ confidence level, the best-fit values are:
$ H_0 = 68.9\pm{2.7} \,~ \mathrm{km\,s^{-1}\, Mpc^{-1}},\quad\Omega_{\rm m} = 0.23\pm{0.14}$ and $
\alpha = 0.047\pm{0.075}$. Furthermore, Figure~\ref{fig2} presents the Hubble data error bars, and Figure~\ref{fig3} shows the residual plot of the Hubble data, compared to the $\Lambda$CDM model with $H_0 = 67.4\,~\mathrm{km\,s^{-1}\,Mpc^{-1}}$ and $\Omega_{\rm m} = 0.315$ \cite{Planck:2018vyg}. This result demonstrates that our framework can accurately describe the Hubble observational data set.

\subsection{Pantheon+ data}

SNeIa are thermonuclear explosions that occur when a white dwarf accretes sufficient mass to exceed the Chandrasekhar limit.
Owing to their remarkably uniform luminosity characteristics, SNeIa serve as excellent standard candles for investigating cosmic dynamics.
Although only a few dozen SNeIa light curves have been recorded by $1980s$, subsequent observational advances have ushered in the modern high-precision era of cosmology.
In this study, we used the most comprehensive supernova sample to date--the Pantheon+ dataset--which incorporates observations from several major international surveys including Pan-STARRS1 (PS1), the Dark Energy Survey (DES), the Supernova Legacy Survey (SNLS), and the Sloan Digital Sky Survey (SDSS) \cite{Riess:2021jrx}.
The complete sample comprises 1,701 SNeIa spectroscopically confirmed subsets spanning a wide redshift range, including low redshift, intermediate redshift, and high redshift subsets \cite{Brout:2022vxf}. Our methodology involves comparing theoretical distance moduli with observed values to constrain cosmological parameters, where the distance modulus for each supernova is calculated using the standard relation
\begin{eqnarray}
\label{10}
\mu_{\mathrm{th}}{\mathrm{(z)}} = 5\log_{10}\left[d_{\rm L}{\mathrm{(z)}}/{\rm Mpc}\right]+25,	
\end{eqnarray}	
where the luminosity distance $d_{\rm L}(z)$ is given by
\begin{equation}
\label{11}
d_{\rm L}(z) = (1+z) \int_0^z \frac{{\rm d}z'}{H(z')}
\end{equation}
The chi-square function $\chi^2$ for Pantheon+ data is
\begin{equation}
\label{12}
\chi_{\rm{SN}}^2 = \sum_{i,j=1}^{1701} \Delta\mu_i \left(C_{\mathrm{SN}}^{-1}\right)_{ij} \Delta\mu_j
\end{equation}	
Here, $C_{\rm SN}$ is the covariance matrix provided in~\cite{SupernovaCosmologyProject:2011ycw}. 
In addition, $\Delta\mu_i$ denotes the difference between the observationally inferred distance 
modulus and the theoretical prediction computed from the model parameters.
\begin{figure}[htbp]
  \centering
  \includegraphics[width=0.6\textwidth]{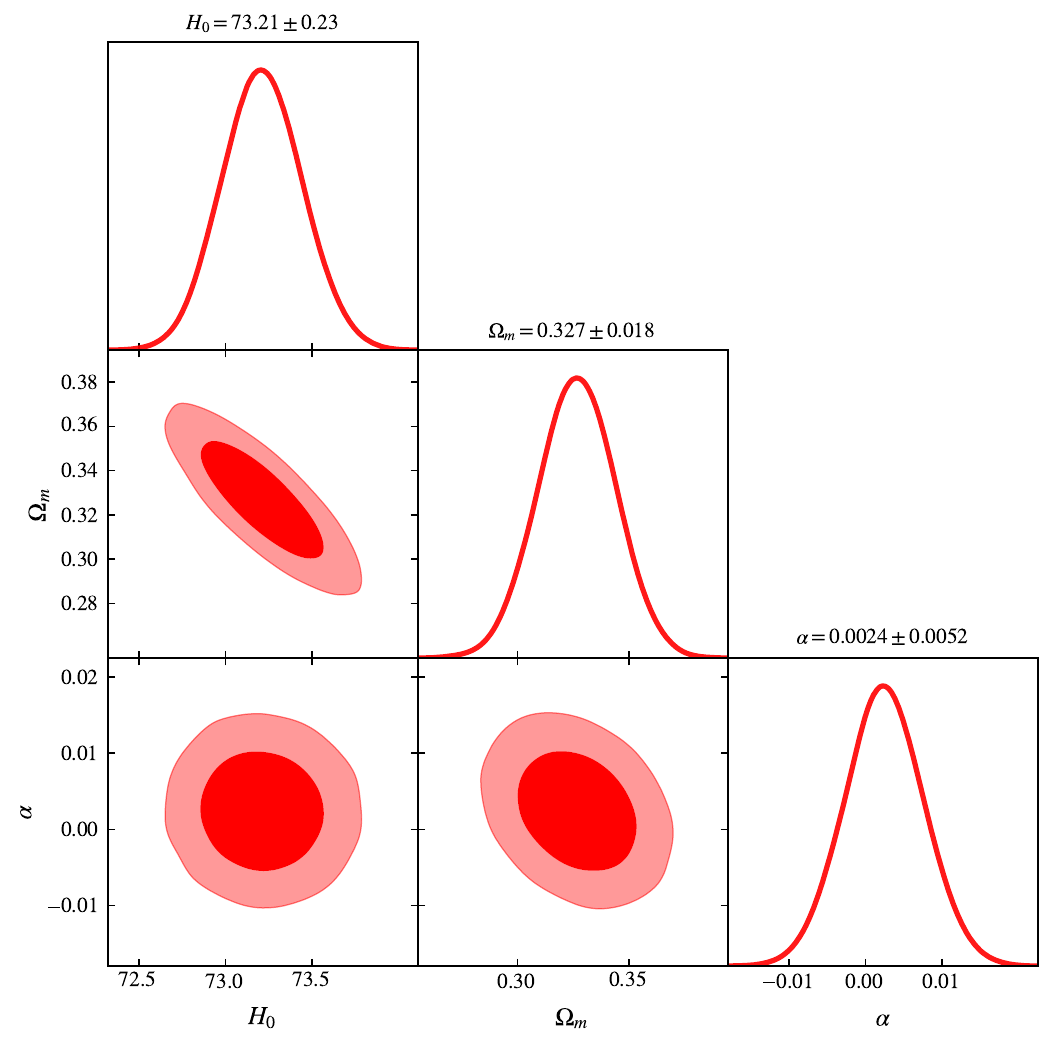}
  \caption{Best-fit parameter estimates together with their $1\sigma$ and $2\sigma$ confidence regions derived from the combined Hubble and Pantheon$+$ observational samples.
  }
  \label{fig4}
\end{figure}
Unlike the previous approach, we now adopt $\chi^2_{\rm H} + \chi^2_{\mathrm{\rm SN}}$ as the minimization constraint for the model parameters $H_0$, $\Omega_{\rm m}$, and $\alpha$. Here, we utilize both the Hubble dataset and Pantheon+ dataset to obtain the best-fit values. As shown in Figure~\ref{fig4}, the results are presented with the confidence level contours $1\sigma$ and $2\sigma$. At $1\sigma$ confidence level, the best-fit values are: $ H_0 = 73.21\pm{0.23} \,~ \mathrm{km\,s^{-1}\, Mpc^{-1}}, \quad
\Omega_{\rm m} = 0.327\pm{0.018}$ and $\alpha = 0.0024\pm{0.0052}$. 

\begin{figure}[htbp]
  \centering
  \includegraphics[width=0.6\textwidth]{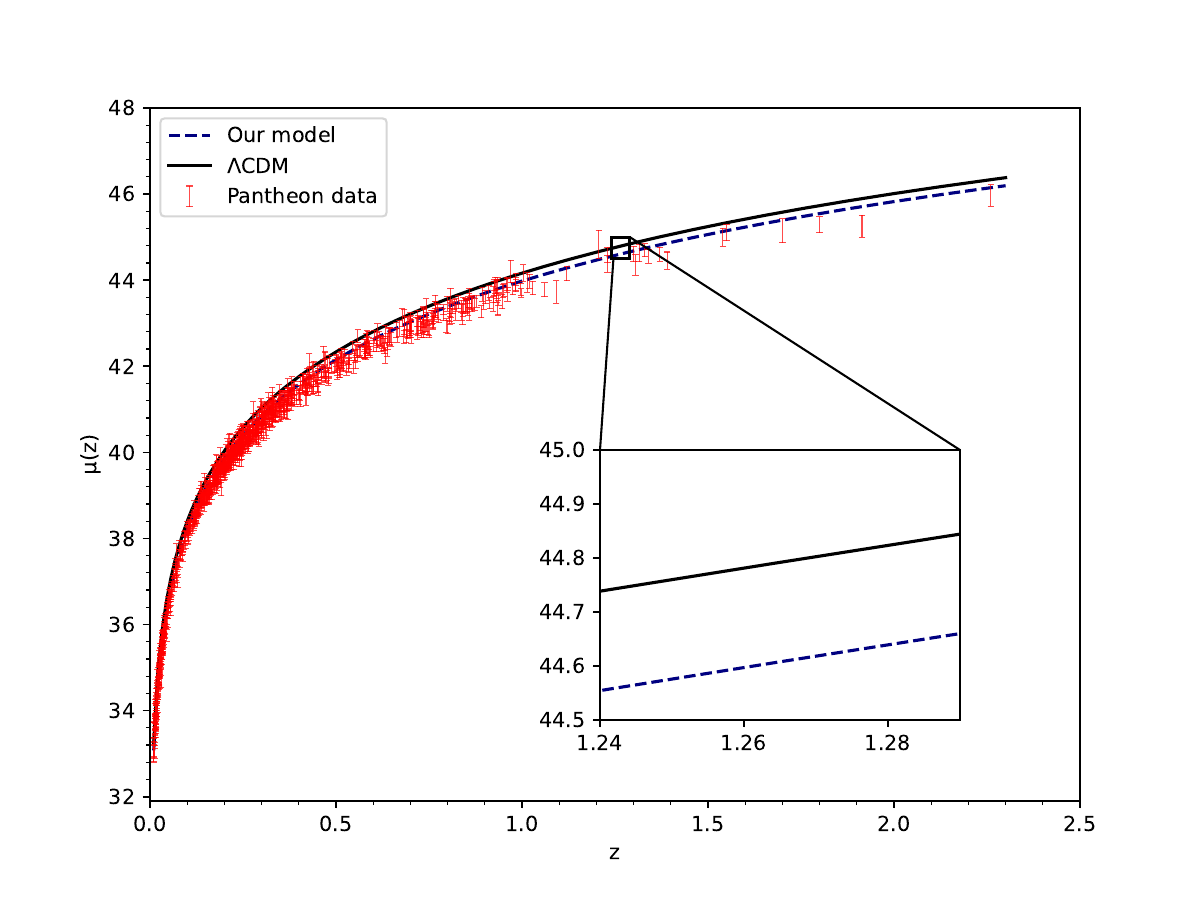}
  \caption{The predicted distance modulus $\mu_{z}$ as a function of redshift $z$ for both our model and the $\Lambda$CDM model, using parameter constraints obtained from the combined Hubble and Pantheon+ data.}
  \label{fig5}
\end{figure}

\begin{figure}[htbp]
  \centering
  \includegraphics[width=0.6\textwidth]{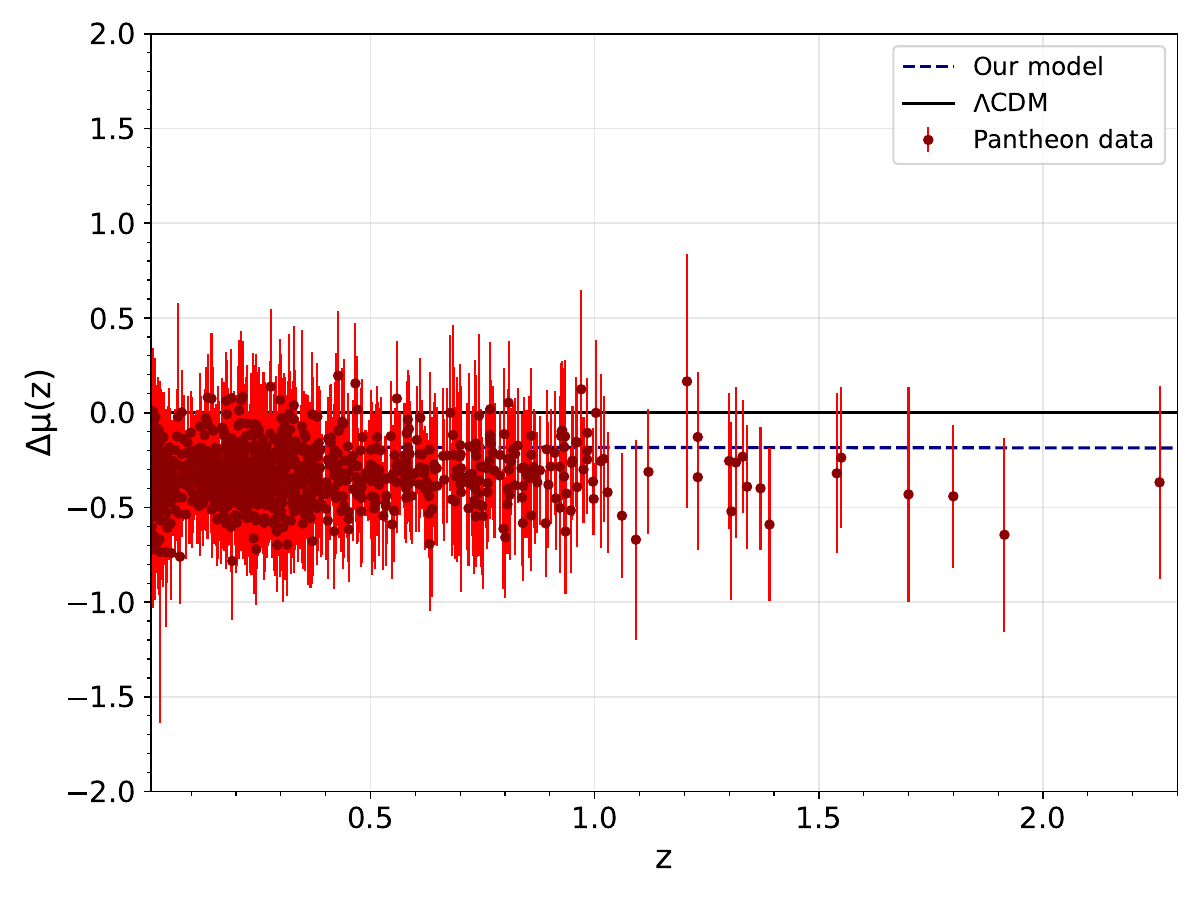}
  \caption{Using the parameter constraints obtained from Hubble + Pantheon+ data, we plot the residuals of the predicted distance modulus $\mu_{z}$ as a function of redshift $z$ to compare our model with the $\Lambda$CDM model.
}
  \label{fig6}
\end{figure}

Additionally, we plot the predicted distance modulus $\mu_{z}$ with the error bar of the supernova data in Figure~\ref{fig5} and the residuals of the predicted distance modulus $\mu_{z}$ to visually display the strengths and weaknesses of our fitting in Figure~\ref{fig6}, where we compare our model with the $\Lambda$CDM model, which shows excellent consistency with the combined Hubble + Pantheon+ dataset.

\subsection{DESI DR2 data}

The most precise BAO measurements to date have been provided by DESI\cite{DESI:2024uvr}, which has observed millions of galaxies and quasars. This unprecedented data set enables the reconstruction of high-fidelity $3D$ maps of the cosmic web, thereby yielding precise determinations of the expansion history of the Universe and imposing stringent constraints on the EoS of dark energy.

\begin{figure}[htbp]
  \centering
  \includegraphics[width=0.6\textwidth]{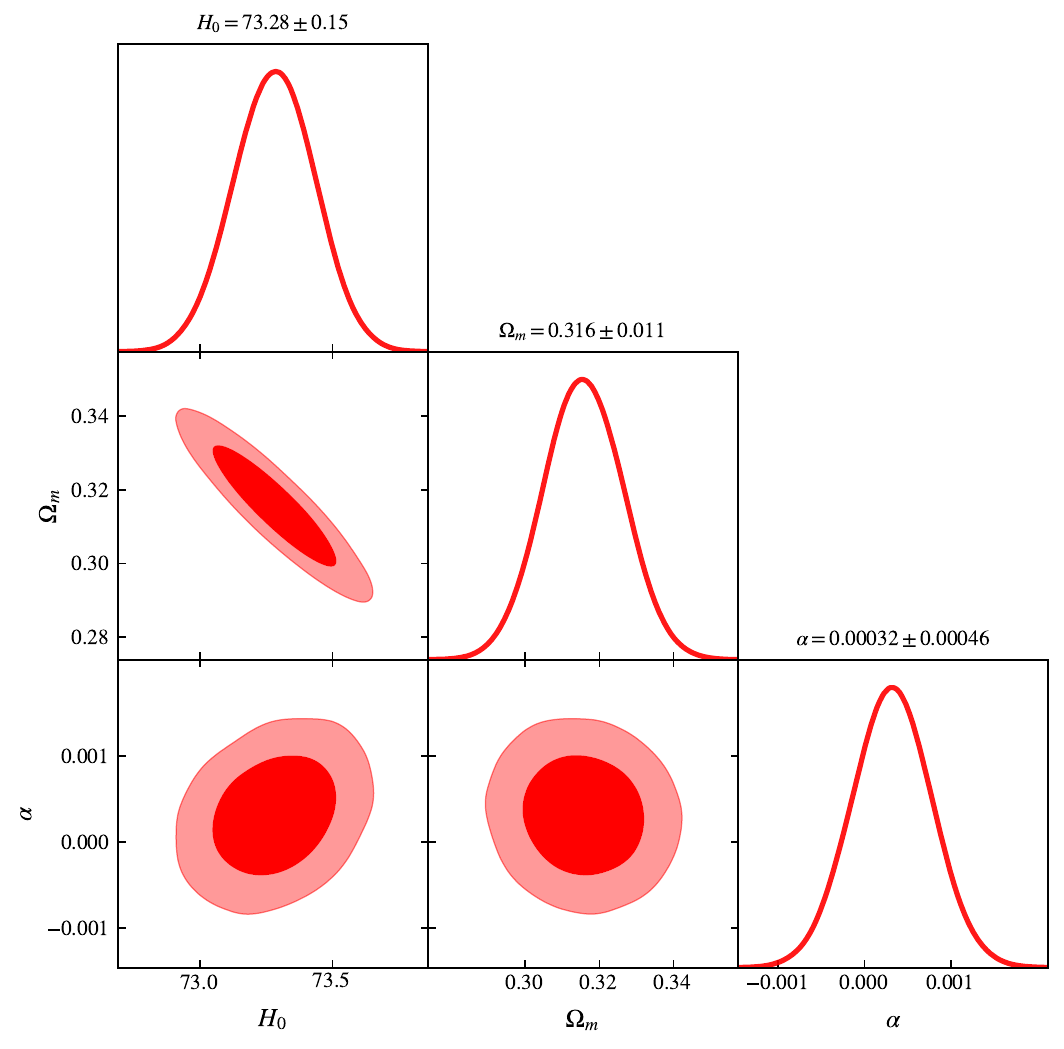}
  \caption{Best-fit constraints on the model parameters with the corresponding $1\sigma$ and $2\sigma$ confidence contours derived from the combined Hubble, Pantheon$+$, and DESI DR2 data.}
  \label{fig7}
\end{figure}

\begin{figure}[htbp]
  \centering
  \includegraphics[width=0.6\textwidth]{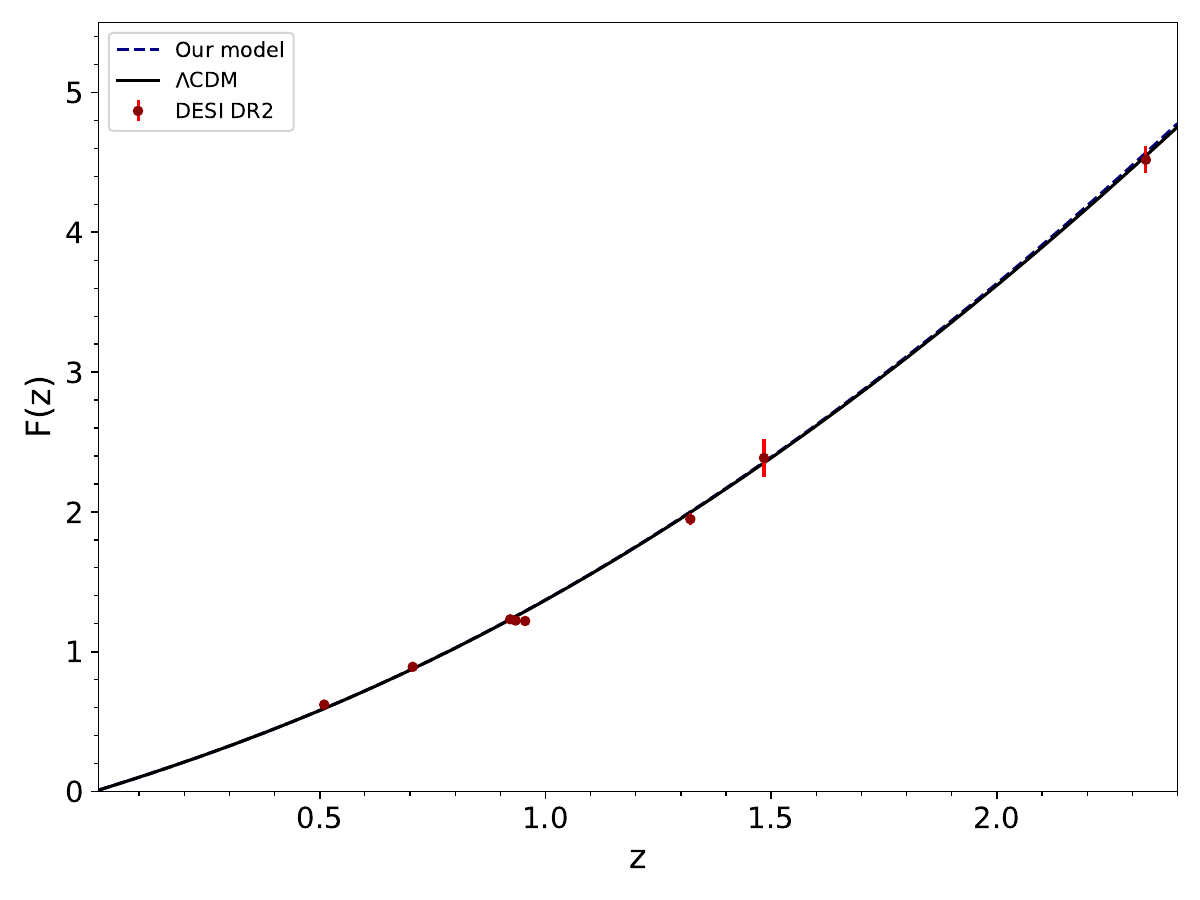}
  \caption{The predicted distance modulus $F_{z}$ as a function of redshift $z$ for both our model and the $\Lambda$CDM model, using parameter constraints obtained from the combined Hubble + Pantheon+ + DESI DR2 data.}
  \label{fig8}
\end{figure}

\begin{figure}[htbp]
  \centering
  \includegraphics[width=0.6\textwidth]{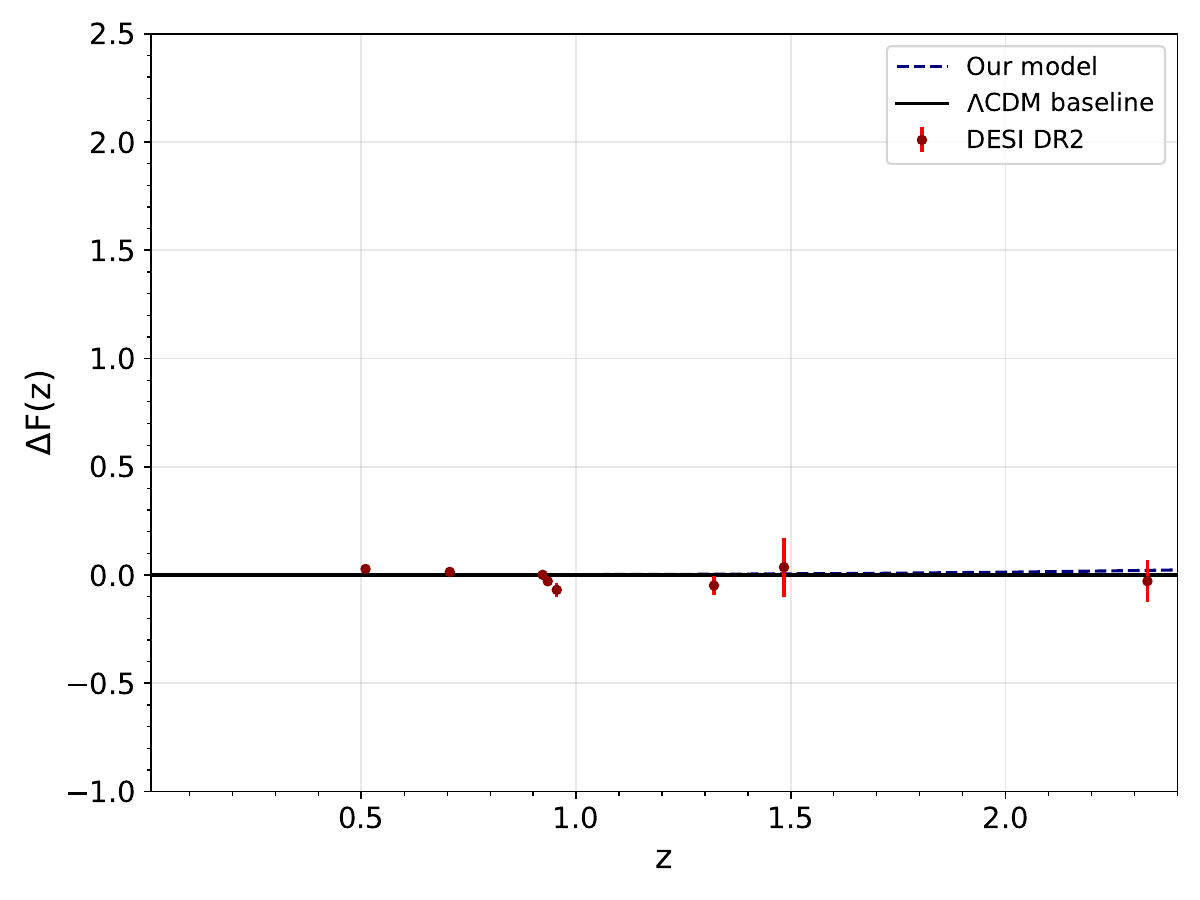}
  \caption{Using the parameter constraints obtained from Hubble + Pantheon$+$ + DESI DR2 data, we plot the residuals of the predicted distance modulus $F_{z}$ as a function of redshift $z$ to compare our model with the $\Lambda$CDM model.}
  \label{fig9}
\end{figure}

The key data set for our analysis is the \( F \) data from DESI DR2 \cite{DESI:2025zgx} (see Table \ref{tab1}), where the measurements are expressed in terms of dimensionless ratios \( D_{\rm M} / r_{\rm d} \), \( D_{\rm V} / r_{\rm d} \), and \( D_{\rm H} / r_{\rm d} \), representing the transverse comoving, angle-averaged, and Hubble distances, respectively, all scaled by the comoving sound horizon \( r_{\rm d} \) at the drag epoch. The function $F$ is defined as $F\equiv D_{\rm M}/ D_{\rm H}$. For a spatially flat universe, the $F$ function takes the form
\begin{equation}
\label{ap}
F=E(z)\int_0^{z}\frac{1}{E(z')}dz',
\end{equation}
where $E(z)=H(z)/H_0$. The chi-square function $\chi^2_{\rm F}$ for the $F$ data takes the following form
\begin{equation}
\label{lum_dist}
\chi_{\rm F}^2 = \sum_{i=1}^{8} \, \frac{[F_i^{\rm th}(H_0,\Omega_{\rm m},\alpha)-F_i^{\rm obs}(z_i)]^2}{\sigma_{\rm F}^2(z_i)},
\end{equation}
where $F_i^{\rm th} $ denotes the theoretical value of the function $F$, while $F_i^{\rm obs}$ represents the observed data, with $\sigma_{\rm F}^{2}$ characterizing the standard error of the observed measurement $F(z)$ at the redshift $z_i$. As illustrated in Figure~\ref{fig7}, we have determined the best-fit values for the model parameters $H_0$, $\Omega_{\rm m}$, and $\alpha$ following the contours $1~ \sigma$ and $2~\sigma$. At $1~ \sigma$ confidence level, the constrained parameters yield: $ H_0 = 73.28\pm{0.15} \,~ \mathrm{km\,s^{-1}\, Mpc^{-1}}, \quad
\Omega_{\rm m} = 0.316\pm{0.011}$ and $\alpha = 0.00032\pm{0.00046}$. 
These results are presented in a comprehensive way in Figure~\ref{fig7}, which displays both the parameter constraints and their correlation properties.
Furthermore, Figure~\ref{fig8} presents the DESI DR2 data error bars, and Figure~\ref{fig9} shows the residual plot of $F$ functions with the DESI DR2 data, compared to the $\Lambda$CDM model with $H_0 = 67.4\,~\mathrm{km\,s^{-1}\,Mpc^{-1}}$ and $\Omega_{\rm m} = 0.315$ \cite{Planck:2018vyg}. This result demonstrates that our framework can accurately describe the combined observational data set.

\begin{table*}
\begin{tabular}{l|l|l}
\hline
index & $z$ & $F$  \\
\hline
$z_1$ & 0.510 & $0.622\pm 0.017$   \\
\hline
$z_2$ & 0.706 & $0.892\pm 0.021$   \\
\hline
$z_3$ &  0.922 & $1.232\pm 0.021$   \\
\hline
$z_4$ &  0.934 & $1.223\pm 0.019$   \\
\hline
$z_5$ &  0.955 & $1.220\pm 0.033$   \\
\hline
$z_6$ &  1.321 & $1.948\pm 0.045$   \\
\hline
$z_7$ & 1.484 & $2.386\pm 0.136$   \\
\hline
$z_8$ & 2.330 & $4.518\pm 0.097$   \\
\hline
\end{tabular}
\caption{$F$ data obtained from BAO measurement.\cite{DESI:2025zgx}}
\label{tab1}
\end{table*}

\section{Results}

\begin{widetext}	
\begin{table*}[!htbp]
\begin{center}
\begin{tabular}{l c c c c c c}
\hline\hline
Data & $H_{0}$ $(\text{km/s/Mpc})$ & $\Omega_{\text{m}}$ & $\alpha$ \\
\hline
Prior   & $(64.9, 76.8)$ & $(0.1, 0.5)$ & $(-0.05, 0.05)$ \\
Hubble & $68.9\pm{2.7}$ & $0.23\pm{0.14}$& $0.047\pm{0.075}$ \\
Hubble + Pantheon$+$   & $73.21\pm{0.23}$ & $0.327\pm{0.018}$ &  $0.0024\pm{0.0052}$ \\
Hubble + Pantheon$+$ + DESI DR2  & $73.28\pm{0.15}$ & $0.316\pm{0.011}$ &  $0.00032\pm{0.00046}$ \\
\hline\hline
\end{tabular}
\caption{A summary of the outcomes inferred from the three datasets.}
\label{tab2}
\end{center}
\end{table*}
\end{widetext}

In this section, we present a comprehensive discussion of the results obtained above. Through MCMC analysis, we have systematically constrained the model parameters ($H_0$, $\Omega_{\rm m}$, and $\alpha$) using progressively combined datasets: (i) Hubble data alone, (ii) Hubble + Pantheon+, and (iii) Hubble + Pantheon+ + DESI DR2. Firstly, the parameters were constrained by using a Hubble data set derived from the cosmic chronometer method. In Figure~\ref{fig1}, the best-fit values of the model parameters were determined, producing a Hubble constant of $ H_0 = 68.9\pm{2.7} \,~ \mathrm{km\,s^{-1}\, Mpc^{-1}}, \quad \Omega_{\rm m} = 0.23\pm{0.14}$ and $\alpha = 0.047\pm{0.075}$, implying that the dark energy behaves like quintessence, but the confidence level is less than 1$\sigma$. The constrained value of $H_0$ is consistent with that obtained from Planck 2018 data, but $\Omega_{\rm m}$ does not \cite{Planck:2018vyg}. Figures~\ref{fig2} and \ref{fig3} show that our model is consistent with both the standard $\Lambda$CDM paradigm and the observational data. 

In order to mitigate the uncertainties in the model parameters, 1701 data points from the Pantheon sample were introduced. Figure~\ref{fig4} presents the joint constraints from the Hubble and Pantheon samples \cite{Perivolaropoulos:2021jda}, giving the best-fit values:
$ H_0 = 73.21\pm{0.23} \,~ \mathrm{km\,s^{-1}\, Mpc^{-1}}, \quad
\Omega_{\rm m} = 0.327\pm{0.018}$ and $\alpha = 0.0024\pm{0.0052}$, indicating that the dark energy also behaves like quintessence, but the confidence level is also less than 1$\sigma$. The constrained value of $\Omega_{\rm m}$ is consistent with that obtained from Planck 2018 data \cite{Planck:2018vyg}, but $H_0$ has a deviation of approximately 4$\sigma$. A significant reduction in the uncertainty of the Hubble parameter was achieved through the joint constraints with the Pantheon compilation. Figures~\ref{fig5} and \ref{fig6} show that our model provides an excellent fit to the Pantheon data while being close to the standard $\Lambda$CDM paradigm. 

Then we included the DESI DR2 data in the analysis and derived the fitting results:
$ H_0 = 73.28\pm{0.15} \,~ \mathrm{km\,s^{-1}\, Mpc^{-1}}, \quad \Omega_{\rm m} = 0.316\pm{0.011}$ and $\alpha = 0.00032\pm{0.00046}$, again suggesting a quintessence dark energy but with the confidence level less than 1$\sigma$. The value of $H_0$ is consistent with that obtained from the Cepheid-calibrated supernova observations \cite{Riess:2021jrx}, but about 4$\sigma$ deviation from the Planck 2018 results \cite{Planck:2018vyg}. Table \ref{tab2} shows that the precision of the constraints in $H_0$, $\Omega_{\rm m}$, and $\alpha$ progressively improves as more observational data are included. Figures~\ref{fig8} and \ref{fig9} show that our model provides an excellent fit to the DESI DR2 while being close to the standard $\Lambda$CDM paradigm. 

To validate our model, we compare its Hubble and Pantheon error bars with the standard $\Lambda$CDM model. Figures~\ref{fig2}, \ref{fig3}, \ref{fig5}, \ref{fig6}, \ref{fig8} and \ref{fig9} demonstrate its excellent concordance with observational datasets. Table \ref{tab2} shows the variations in the best-fit parameters ($H_0$, $\Omega_{\rm m}$, $\alpha$) depending on the data set used. The OHD only fit deviates significantly from the results of the joint analyses, while the two joint analyses are mutually consistent. This indicates that the Hubble dataset, likely due to its limited volume, can yield biased constraints when used independently. Since $\alpha$ is statistically indistinguishable from zero, our model may be difficult to distinguish from $\Lambda$CDM with observations.

For model comparison, we employ the Akaike information criterion (AIC) 
and the Bayesian information criterion (BIC), which are defined, respectively, as
\begin{equation}
\mathrm{AIC} = -2\ln\mathcal{L}_{\max} + 2k,\qquad
\mathrm{BIC} = -2\ln\mathcal{L}_{\max} + k\ln N,
\end{equation}
where $k$ is the number of free parameters and $N$ is the number of 
data points\cite{Liddle:2007fy}.
Finally, we statistically compare our model with the $\Lambda$CDM and $\omega$CDM models. For the $\Lambda$CDM model, we find $\Delta \mathrm{AIC} = 1.17$ and $\Delta \mathrm{BIC} = 7.23$. Since $\Delta \mathrm{AIC} < 2$, the two models exhibit an almost indistinguishable goodness of fit, while $\Delta \mathrm{BIC} = 7.23$ provides moderate evidence in favor of $\Lambda$CDM, mainly due to the heavier penalty imposed by the BIC on the additional parameter $\alpha$. In general, $\Lambda$CDM is statistically slightly preferred, but
our model still fits the data well.
Compared with the $\omega$CDM model, we obtain $\Delta \mathrm{AIC} = \Delta \mathrm{BIC} = 3.67$. Within the framework of the interpretation of the information criteria, $\Delta = 3.67$ falls in the range of $2 < \Delta < 4$, indicating moderate support for the $\omega$CDM model over our parameterization. Although $\omega$CDM shows a discernible statistical advantage, the difference is not decisive. 
Nevertheless, further analysis reveals that our model yields a Hubble constant value of $H_0 = 73.28 \pm 0.15\ \mathrm{km\ s^{-1}\ Mpc^{-1}}$, which agrees remarkably well with the late-universe measurement from the SH0ES collaboration and differs from the early-universe Planck result by about $5\sigma$. This shows that our model can effectively alleviate the Hubble tension, a potential advantage worth further research.

\section{Conclusion}

We have proposed a parameterized EoS for dark energy, which was constrained through MCMC using a combination of Hubble, Pantheon+, and DESI DR2 data.

To minimize the impact of systematic uncertainties and any potential biases inherent in the observational data, we adopted the following methodology: we performed cross-validation of the constraints using multiple independent datasets to mitigate the influence of systematic errors from any single dataset. Furthermore, in selecting the datasets for our constraints, we specifically employed those that have undergone extensive cross-calibration and systematic error correction by other researchers. In addition, we conducted a comprehensive error propagation analysis incorporating error terms into the Monte Carlo likelihood function. Despite implementing extensive measures to minimize uncertainties throughout our analysis and simulations, the complete elimination of biases and systematic errors from observational data remains challenging and ultimately unattainable.

The model was further validated against Hubble, Pantheon+, and DESI DR2 datasets, and the values of the best-fit parameters are constrained as $ H_0 = 73.28\pm{0.15} \,~ \mathrm{km\,s^{-1}\, Mpc^{-1}}$, \quad $\Omega_{\rm m} = 0.316\pm 0.011$ and $\alpha = 0.00032\pm{0.00046}$,
indicating that the dark energy behaves like quintessence with a confidence level of less than 1$\sigma$, meaning that it is unlikely to be distinguished from $\Lambda$CDM. As provided in Table \ref{tab2}, we have reported the upper and lower bounds for each parameter. It is clearly evident that the results demonstrate minimal sensitivity to these errors. Furthermore, the confidence intervals also show that these uncertainties are small and do not substantially influence the conclusions reached in this work.

These results show excellent agreement with Cepheid-calibrated supernova observations, but exhibit a significant discrepancy (exceeding reported uncertainties) compared to the 2018 Planck CMB measurements.  Considering both the statistical performance and the ability to relieve the Hubble tension, this single-parameter parameterization remains a noteworthy cosmological model. If future high-precision sky surveys provide more data, such as the Dark Energy Survey (DES) or the Kilo-Degree Survey (KiDS), our research can be further expanded.

\section*{Acknowledgments}
This study is supported in part by the National Natural Science Foundation of China (Grant No. 12333008) and the Yunnan Fundamental Research Projects (YFRP) grant No. 202501CF070016
\bibliographystyle{elsarticle-num}
\bibliography{reff}
\end{document}